\documentclass[epj]{svjour}
\usepackage{graphics}

\newcommand{\rmi}{\mathrm{i}}
\newcommand{\rmd}{\mathrm{d}}
\newcommand{\psin}[1]{\psi_{\gamma_{#1}}}
\newcommand{\mn}[1]{m_{\gamma_{{#1}}}}
\newcommand{\xivn}[1]{\xi_{v_{#1}}}
\newcommand{\xin}[1]{\xi_{\gamma_{#1}}}

\newcommand{\TM}[1]{\mathbf{M}^{(\gamma_n)}}\sffamily
\newcommand{\GMo}[1]{\mathbf{\Gamma}^{(1,\gamma_{#1})}}
\newcommand{\GMt}[1]{\mathbf{\Gamma}^{(2,\gamma_{#1})}}
\newcommand{\GMoel}[2]{\mathrm{\Gamma}^{(1,\gamma_{#1})}_{#2}}
\newcommand{\GMtel}[2]{\mathrm{\Gamma}^{(2,\gamma_{#1})}_{#2}}

\newcommand{\sen}[2]{#1_{\gamma_{#2}}}

\newcommand{\lf}{l_{\gamma_1}}
\newcommand{\ls}{l_{\gamma_2}}

\begin{document}

\title{On the electron scattering on the one-dimensional complexes: the vertex amplitudes method}
\titlerunning{On the electron scattering in one-dimensional complexes: the vertex amplitudes method}
\author{Alexander F.˜Klinskikh\thanks{\emph{e-mail:} klinskikh@live.ru} \and Anton V.˜Dolgikh\thanks{\emph{e-mail:} dolgikh@niif.vsu.ru} \and Peter A.˜Meleshenko\thanks{\emph{e-mail:} melechp@yandex.ru} \and Sergey A. Sviridov \and Hang T.T. Nguyen
}
\authorrunning{A.F. Klinskikh \textit{et al.}}
\institute{Theoretical Physics Department, Voronezh State University, Universitetskaya sq.1, 394006 Voronezh, Russia}

\date{Received: date / Revised version: date}

\abstract{
The problem of electron scattering on the one-dimensional complexes is considered. We propose a novel theoretical approach to solution of the transport problem for a quantum graph. In the frame of the developed approach the solution of the transport problem is equivalent to the solution of a linear system of equations for the \emph{vertex amplitudes} $\mathbf{\Psi}$. All major properties, such as transmission and reflection amplitudes, wave function on the graph, probability current, are expressed in terms of one $\mathbf{\Gamma}$-matrix that determines the transport through the graph. The transmission resonances are analyzed in detail and comparative analysis with known results is carried out.
}

\maketitle

\section{Introduction}
Impressive progress in the development, manufacturing and understanding of various nanostructures is observed in recent times~\cite{Umbach84,Umbach86,Webb85,Chandrasekhar85,Levy90,Timp87,Dolan86,Kurdak92,Benoit86,Ishibashi87}. This makes scientists not only to formulate the new theoretical problems and to reconsider the existing ones but also to look for the new approaches to the solution of these problems. That ''new'' approaches usually have their own history.

The classical example is the dynamics of the electron in a graph (one-dimensional complex~\cite{Bamberg91}). This problem has more than seventy years history and was introduced into physics in connection with calculation of the properties of conjugated organic molecules~\cite{Pauling36,Ruedenberg1953,Platt49_1,Platt49_2,Schmidtke66,Mallion75}, but
then gradually lost its relevance to the quantum chemistry. Nevertheless, the model of electron in a graph stimulated the emergence of the new field in the mathematics related to the investigation of the spectrum of self-adjoint operators in compact and noncompact graphs~\cite{Dowker72,Schulman71,Budgor76,Ringwood81,Gerasimenko88_1,Gerasimenko88_2}.  In recent years the renewed interest in this area involved significant redevelopment and enhancement of the existing theory~\cite{Texier2001,Desbois2000a,Akkermans2000}.

The quantum mechanics in the graph demonstrates many interesting peculiarities. One of them is the spectrum of the Schr\"{o}dinger operator. To find it one should construct its self-adjoint extension which is not unique~\cite{Schulman71}. This is most commonly done in way that was first proposed by Griffith~\cite{Griffith53}.
The classical counterpart of the Griffith's extension is the well known Kirchhoff rules for current conservation in each node supplemented with the wave function continuity condition resulting from the quantum mechanics ideology. The spectrum of this self-adjoint extension is discrete on compact graph~\cite{Gerasimenko88_1,Kuchment04}.

The more interesting property of the spectrum is the existence of the discrete eigenenergies embedded in the continuum (in what follows, it is abbreviated as BIC). The possibility of such states was shown in the prominent paper of Wigner and von Neumann~\cite{Wigner1929} and the method to construct them was proposed. Later, it was improved by Stillinger and Herric~\cite{Stillinger75}. The central idea of this approach consists in such a choice of the potential that the continuum wave function becomes integrable. In the same work of Stillinger and Herrick the BIC were predicted in the semiconductor superlattices. In 1992 it was confirmed in the experiment on the IR irradiation of the $\mathrm{Al_{0.48}In_{0.52}As/Ga_{0.47}In_{0.53}As}$ superlattice~\cite{Capasso92}.

The discrete spectrum superimposed on the continuum in quantum graph was considered in details in a number of papers~\cite{Texier2001,Sadreev06,Texier2003}. Such states in the graph behave quite similarly to the ordinary (in the presence of potential) BIC. This fact encouraged Bulgakov~\textit{et.al.}~\cite{Sadreev06} to use the term BIC in the case of graph. In the works of Texier~\cite{Texier2001,Texier2003} the presence of the discrete eigenvalues for positive energies on the graph is more carefully interpreted as decoupling of the compact part of the graph from the external leads.

The considered phenomena can be interpreted in the formalism of the de Broglie waves dynamics~\cite{Xia92}. However the results of Xia~\cite{Xia92} are rather cumbersome to analyze and to make any generalizations.

In connection with the above the aim of the presented work consists in the development of the successive mathematical technique based on the scattering theory to solve the following problems: wave function of the electron in the one-dimensional complexes (graph-like structures), energy spectrum, transmission and reflection amplitudes, Green's function and response to the external electromagnetic field. These objects are treated in the frame of the single technique in contrast to the current papers in this field that use the different mathematical methods. In addition, the developing approach gives the results with clear physical meaning. Moreover, using the proposed technique allows not only to explain the some fascinating effects~\cite{Texier2001,Capasso92,Sadreev06}, but also to formulate the conditions of their occurrence.

This paper is organized as follows. In section~\ref{sec:Preliminaries} we briefly consider the basics of electron transmission through the quantum graphs; we give the general considerations and equations for electron moving on the quantum graph.
The main mathematical formalism of the developed method and its application to the electron transmission through the potential with finite support is described in section~\ref{sec:Main formalism}. In section~\ref{sec:Scattering by the quantum ring} we apply the proposed formalism to the problem of electron transmission through symmetric and asymmetric (with and without Aharonov-Bohm flux) quantum ring. In section~\ref{sec:Conclusions} the obtained results are summarized. The application of the \emph{vertex amplitudes} method to the classical quantum mechanics problem of electron scattering by the potential with finite support is presented in Appendix~\ref{app:potential}. In Appendix~\ref{app:parallel} the problem of electron scattering by the system of parallel coupled quantum wells is considered.

\section{Preliminaries\label{sec:Preliminaries}}
\begin{figure}[h]
\begin{center}
\resizebox{0.45\textwidth}{!}{%
  \includegraphics{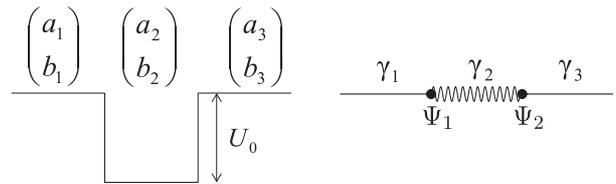}
}
\caption{The one-dimensional scattering on the potential well (barrier) as a problem on the graph. Wavy line represents the potential well (barrier), $\mathrm{\Psi}_1$ and $\mathrm{\Psi}_2$ are the values of the electron wave function at the graph's vertices, $\gamma_1$, $\gamma_2$ and $\gamma_3$ are the graph's edges. The coefficients $(a_i,b_i)$, $i=1,2,3$ are the coefficients of the wave function: $\psi_i(x)=a_ie^{\rmi kx}+b_ie^{-\rmi kx}$.}
\label{fig: quantum well and quantum graph}
\end{center}
\end{figure}

Most of the problems of one-dimensional quantum mechanics can be formulated in the language of quantum graphs. Moreover, the solution of these problems in the frame of the graphs formalism could be more efficient and ''elegant'' in comparison with the traditional methods of quantum mechanics. One of such problems is the electron scattering on the system of connecteed in parallel quantum wells (the detailed analysis is carried out in Appendix~\ref{app:parallel}). A simple and clear example which demonstrates the use of the graphs formalism is the problem of the one-dimensional quantum well (see fig.~\ref{fig: quantum well and quantum graph}).

There are various ways to solve the Schr\"{o}dinger equation on the graph~\cite{Xia92,Texier2001,Akkermans2000}. In the presented work we develop the \emph{vertex amplitudes} method. The main idea of this method is to express the parameters of the problem through the values of the wave function at the vertex of graph $\mathrm{\Psi}_1$ and $\mathrm{\Psi}_2$. As will be shown below the considered method allows to solve the scattering problem in the quantum graph, to find the energy spectrum and to construct the wave functions. The classical analog of the presented \emph{vertex amplitudes} method is the Kirchhoff's nodal potentials method in the theory of electrical circuits.

Let us consider the one-dimensional scattering problem for a quantum graph with the potentials that are placed on the graph's edges. The quantum graph is a metric graph with the self-adjoint Hamilton operator defined on it~\cite{Kuchment08}.

We assume that the quasi-one-dimensional dynamics takes place. The quasi-one-dimensional dynamics may be realized in the lowest sub-band of a very narrow quantum wire. The wire width in the real experiments can not be infinitely narrow. But, for the quantum wire under low temperature the electron dynamics is quasi-one-dimensional because the higher transverse levels can not be excited.

The considered quantum graph consists of a compact part connected with the reservoirs of the charge carries by the semi-infinite edges. We denote these edges as the \textit{in,out}-edges. In this edges the asymptotic conditions for the electron wave functions with respect to the compact part of the graph are realized.

\begin{figure}[h]
\begin{center}
\resizebox{0.48\textwidth}{!}{%
  \includegraphics{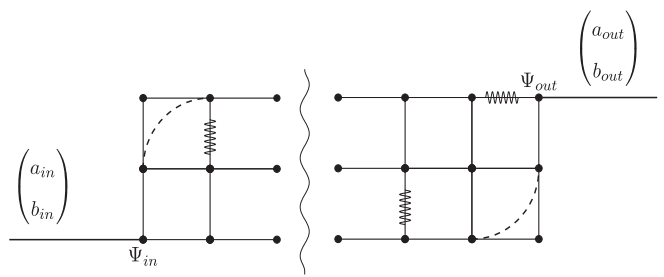}
}
\caption{ General form of a quantum graph considered in the paper. Coefficients $(a_{in},\,b_{in})$, $(a_{out},\,b_{out})$ correspond to the \textit{in,out}-wave functions (see equations~(\ref{eq:psi in},~\ref{eq:psi out}). Wavy lines represent the potentials on the edges. Dashed lines show that the node with only two intersected edges equivalent to the one edge without nodes.}
\label{fig:general graph}
\end{center}
\end{figure}

In the general case, topology of the graph may be fairly sophisticated. In this connection the graph considered here is the two-dimensional array with the real potentials with finite support placed on the edges (see fig.~\ref{fig:general graph}).

The wave functions of the electron in the \textit{in,out}-edges are:
\begin{equation}\label{eq:psi in}
\psi_{in}=a_{in}\exp(\rmi k x) + b_{in}\exp(-\rmi k x),
\end{equation}
\begin{equation}\label{eq:psi out}
\psi_{out}=a_{out}\exp(\rmi k x) + b_{out}\exp(-\rmi k x),
\end{equation}
where $k$ is the electron wave-number. As it follows from~(\ref{eq:psi in}) and~(\ref{eq:psi out}) the elastic scattering takes place. This assumption is not necessary and does not influence on the generality of the proposed method, but facilitates further mathematics. The physical meaning of the coefficients $a_{in}$, $b_{in}$, $a_{out}$, $b_{out}$ depends on the considered problem (see tab.~\ref{tab:coefficients for different scattering problems}).

\begin{table}[h]
\caption{\label{tab:coefficients for different scattering problems} Coefficients of the wave functions for different ``scattering problems'' (``sc.pr.'').}
\begin{center}
\begin{tabular}{ll}
\textbf{``Left sc. pr.''} & \textbf{First Jost's solution}\\
\hline\hline\noalign{\smallskip}
$a_{in}(k)=1$   &   $a_{in}(k)=\mathrm{M}_{11}$ \\
$b_{in}(k)=r(k)$ & $b_{in}(k)=\mathrm{M}_{21}$\\
\noalign{\smallskip}\hline
$a_{out}(k)=t(k)$ & $a_{out}(k)=1$,\\
$b_{out}(k)=0$ &   $b_{out}(k)=0$\\
\hline\hline\noalign{\smallskip}
\textbf{Second Jost's solution} & \textbf{``Right sc. pr.''}\\
\hline
$a_{in}(k)=0$ &  $a_{in}(k)=0$\\
$b_{in}(k)=1$ & $b_{in}(k)=t(k)$\\
\noalign{\smallskip}\hline
$a_{out}(k)=-\mathrm{M}_{12}$ & $a_{out}(k)=r(k)$\\
$b_{out}(k)=\mathrm{M}_{11}$  & $b_{out}(k)=1$\\
\noalign{\smallskip}\hline
\end{tabular}
\end{center}
\end{table}
In table~\ref{tab:coefficients for different scattering problems} $\mathbf{M}$ represents the transfer matrix~\cite{Merz70,Klinskikh08} and $r(k)$, $t(k)$ are the reflection and transmission amplitudes respectively~\cite{LandauIII,Novikov84}. The explicit form of $\mathbf{M}$ is discussed below.

In each edges $\gamma_n$ of length $l_{\gamma_n}$ the proper coordinates are used $\xin{n}\in [0,l_{\gamma_n}]$. In the \textit{in,out}-edges coordinates are defined in a different way: $\xi_{in} \in (-\infty, 0)$, $\xi_{out}\in(0,\infty)$. These coordinates are just a natural parameter in differential geometry~\cite{Dubrovin92}. The electron wave function $\mathrm{\Psi}$ in the graph is represented by a set of components $\psin{n}(\xin{n})$, $n=1,2,\ldots,M$, where $M$ is the number of edges in graph. Each element $\psin{n}(\xin{n})$ of the set is governed by the Schr\"{o}dinger equation:
\begin{equation}\label{eq:Schroedinger equation on graph}
\begin{array}{c}
H_{\gamma_n}\psin{n}(\xin{n})=\varepsilon_{\gamma_n}\psin{n}(\xin{n}),\\\\
H_{\gamma_n}=\displaystyle\left[-\frac{\hbar^{2}}{2}\frac{d}{d\xin{n}}\left(\frac{1}{\mn{n}}\frac{d}{d\xin{n}}\right)+V_{\gamma_n}(\xin{n})\right], \\\\
D(H_{\gamma_n})=\left\{ \psi_{\gamma_n}: \psi_{\gamma_n}\in C_0^{\infty}[\gamma_n] \right\},
\end{array}
\end{equation}
where $\mn{n}$ is an effective mass of the electron in the edge $\gamma_n$, $V_{\gamma_n}$ is the real and locally measurable potential placed on the edge $\gamma_n$. The operators $H_{\gamma_n}$ are the symmetric operators with the defect indexes $(2,2)$~\cite{Gerasimenko88_1}. It is known~\cite{Richtmyer,Akhiezer}, that there exists a self-adjoint extension $H_{\mathcal{G}}$ of operator
$$
\widetilde{H}_{\mathcal{G}}=\sum\limits_{i=n}^{M} \bigoplus H_{\gamma_n},
$$
where $M$ is a number of edges in the graph $\mathcal{G}$. The self-adjoint extension $H_{\mathcal{G}}$ can be constructed by imposing additional boundary conditions on the wave functions $\psi_{\gamma_n}$~\cite{Richtmyer,Akhiezer}, namely:
\begin{enumerate}
\item the standard continuity conditions at vertices $v_{n}$:
\begin{equation}\label{eq:continuity}
\psin{1}(\xivn{n})=\ldots=\psin{N}(\xivn{n})=\Psi_n,
\end{equation}
where $N$ is a number of edges intersected at vertex $v_n$, $\xivn{n}$ is the value of local coordinate $\xin{n}$ in vertex $v_n$, and  $\Psi_n\equiv \Psi(\xivn{n})$;
\item the restriction on the sum of derivatives of the wave functions at vertex $v_n$:
\begin{equation}\label{eq:hermiticity}
\displaystyle{\sum\limits_{j=1}^{N}\varepsilon_j\frac{1}{\mn{j}}\frac{\rmd\psin{j}}{\rmd\xi_{\gamma_j}}\Bigg|_{\xi_{\gamma_j}=\xivn{n}}=0},
\end{equation}
where
$$\varepsilon_j=\left\{\begin{array}{cl}
-1, & \mbox{for the edges terminating at vertex $v_n$}\\
                     \phantom{-}1, & \mbox{for the edges emmanating from vertex $v_n$}\end{array}\right.$$
and the index $j$ numerates the edges which are incoming at vertex $v_n$ or outgoing from it.
\end{enumerate}
Let us note that the boundary conditions~(\ref{eq:hermiticity}) are similar to the well known Kirchhoff rules~\cite{Akkermans2000,Nazarov09}.

In what follows, we work with the self-adjoint operator $\widetilde{H}_{\Gamma}$ only and the tilde is omitted for simplicity.

Now, we consider the edge $\gamma_n$ with an potential that is localized somewhere on this edge. The wave function $\psin{n}$ can be determined by the following pairs of coefficients: $(\sen{a}{n},\sen{b}{n})$ and $(\sen{c}{n},\sen{d}{n})$. They have the following meaning: the wave function in every edge outside the potential can be taken in the form $a\exp(\rmi k x)+b\exp(-\rmi k x)$, thereby, the pair $(\sen{a}{n},\sen{b}{n})$ are the coefficients of $\psin{n}$ before the potential and the pair $(\sen{c}{n},\sen{d}{n})$ are the coefficients of $\psin{n}$ beyond the potential.

It should be noted that the wave functions in the edge is not necessary to be a linear combination of the plane waves $\exp(\pm\rmi k x)$. If there any potential on the edge that nowhere turns into zero then the wave function is a linear combination of two appropriate solutions of Schr\"{o}dinger equation. Such potentials, (\textit{e.g.} $V(x)=-V_0\,\mathrm{sech}^2\gamma x$) were used for description of the aromatic molecules in the frame of the ``graph model'' (when the aromatic molecule can be treated as a graph)~\cite{Montroll70}.

\section{Main formalism\label{sec:Main formalism}}
Let us consider the left scattering problem.  Thus, according to table~\ref{tab:coefficients for different scattering problems}, the \textit{in,out}-wave functions are:
\begin{equation}\label{eq:psi-in,out for left scattering}
\begin{array}{lll}
\psi_{in}&=&\exp(\rmi k x)+r(k)\exp(-\rmi k x),\\
\psi_{out}&=&t(k)\exp(\rmi k x).
\end{array}
\end{equation}

An application of the conditions~(\ref{eq:continuity}),~(\ref{eq:hermiticity}) leads to the system of linear algebraic equations for the \emph{vertex amplitudes} $\mathbf{\Psi}=(\mathrm{\Psi}_{in}\equiv\mathrm{\Psi}_1,\,\mathrm{\Psi}_2,\ldots,\mathrm{\Psi}_{out}\equiv\mathrm{\Psi}_n)^T$. This system contains the coefficients of functions $\psin{n}$. Hence to solve the obtained system for $\mathbf{\Psi}$ it is needed to express $a_{\gamma_n}$, $b_{\gamma_n}$, $c_{\gamma_n}$ and $d_{\gamma_n}$ in terms of $(\mathrm{\Psi}_1,\,\mathrm{\Psi}_2,\ldots,\mathrm{\Psi}_n)$. To do this we use the following obvious relations:
\begin{equation}\label{eq:gamma matrices}
\left(\begin{array}{lcr}
\sen{a}{n} \\ \sen{b}{n} \end{array} \right)=\GMo{n}\left(\begin{array}{lcr}
\mathrm{\Psi}_n\\ \mathrm{\Psi}_{n+1} \end{array} \right), \left(\begin{array}{lcr}
\sen{c}{n} \\ \sen{d}{n} \end{array} \right)=\GMt{n}\left(\begin{array}{lcr}
\mathrm{\Psi}_n\\ \mathrm{\Psi}_{n+1} \end{array} \right).
\end{equation}
The pairs of coefficients $(\sen{a}{n},\sen{b}{n})$ and $(\sen{c}{n},\sen{d}{n})$ are related by the transfer-matrix $\mathbf{M}$~\cite{Merz70,Klinskikh08}:
\begin{equation}\label{eq:transfer matrix}
\quad \left(\begin{array}{lcr}
\sen{a}{n} \\ \sen{b}{n} \end{array} \right)=\TM{n}\left(\begin{array}{lcr}
\sen{c}{n}\\ \sen{d}{n} \end{array} \right).
\end{equation}
The elements of the M-matrix are
\begin{equation}\label{eq:general form of transfer matrix}
\mathbf{M}^{(\gamma_n)}=\left(
\begin{array}{lcr}
1/t_{\gamma_n} & r_{\gamma_n}/t_{\gamma_n}\\
r^*_{\gamma_n}/t^*_{\gamma_n} & 1/t^*_{\gamma_n}
\end{array}
\right),
\end{equation}
$t_{\gamma_n}(r_{\gamma_n})$ is the transmission (reflection) amplitude for the potential placed on the edge $\gamma_n$.
Using relations~(\ref{eq:gamma matrices}),~(\ref{eq:transfer matrix}) and condition~(\ref{eq:hermiticity}) one get:
\begin{equation}\label{eq:gamma matrices explicit form}
\begin{array}{l}
\mathbf{\Gamma}^{(1,\gamma_n)}=\displaystyle{\frac{1}{\zeta-\zeta^{*}}}
\left(\begin{array}{cc}
\zeta&-1 \\
-\zeta^{*}&1
\end{array}\right), \mbox{where}\\\\
\zeta=\exp(-\rmi k \xivn{n+1})\mathrm{M}^{(\gamma_n)}_{11} - \exp(\rmi k \xivn{n+1})\mathrm{M}^{(\gamma_n)}_{12}.
\end{array}
\end{equation}
From~(\ref{eq:gamma matrices}),~(\ref{eq:transfer matrix}) it follows
\begin{equation}\label{eq:second gamma matrix}
\GMt{n}=\big[\TM{n}\big]^{-1}\GMo{n}.
\end{equation}
Finally,~(\ref{eq:hermiticity}),~(\ref{eq:gamma matrices}) and~(\ref{eq:gamma matrices explicit form}),~(\ref{eq:second gamma matrix}) lead to the system of linear equation for unknown vector $\mathbf{\Psi}$. Finding the $\mathbf{\Psi}$ solves the transport problem for graph because the wave functions $\psi_{in}$, $\psi_{out}$ satisfy the following conditions at the \textit{in,out}-vertices:
\begin{equation}\label{eq:conditions at in out vertices}
1+r(k)=\mathrm{\Psi}_{in}, \quad t(k)=\mathrm{\Psi}_{out}.
\end{equation}

The wave function on the edge can be constructed using~(\ref{eq:gamma matrices}),~(\ref{eq:gamma matrices explicit form}). If there is no potential on the edge then $\GMo{1}$ and $\GMt{2}$ are identical and:
$$
\begin{array}{l}
a_{\gamma_n}=\GMoel{n}{11}\mathrm{\Psi}_n+\GMoel{n}{12}\mathrm{\Psi}_{n+1}\,,\\\\
b_{\gamma_n}=\GMoel{n}{21}\mathrm{\Psi}_n+\GMoel{n}{22}\mathrm{\Psi}_{n+1}\,,
\end{array}
$$
or
$$
\begin{array}{l}
a_{\gamma_n}=\displaystyle{\frac{\rmi}{2\sin kl_{\gamma_n}}\left[\exp(-\rmi kl_{\gamma_n})\mathrm{\Psi}_n-\mathrm{\Psi}_{n+1}\right]}\,,\\\\  b_{\gamma_n}=\displaystyle{\frac{\rmi}{2\sin kl_{\gamma_n}}\left[\mathrm{\Psi}_{n+1}-\exp(\rmi kl_{\gamma_n})\mathrm{\Psi}_n\right]}\,.
\end{array}
$$
The last gives
\begin{equation}\label{eq:general form of wavefunction}
\psin{n}=\displaystyle{\frac{1}{\sin k l_{\gamma_n}}}\left[\mathrm{\Psi}_{n+1}\sin k\xin{n}-\mathrm{\Psi}_n \sin(k\xin{n}-k l_{\gamma_n})\right].
\end{equation}
The obtained form of the wave function~(\ref{eq:general form of wavefunction}) still correct for a general graph without potential on the edges and is used in a number of works~\cite{Texier2001,Akkermans2000,Alexander83}.

The wave function~(\ref{eq:general form of wavefunction}) creates the current $j_{\gamma_n}$ in the edge $\gamma_n$:
\begin{equation}\label{eq:current on the edge}
\begin{array}{c}
j_{\gamma_n}= k |\mathrm{\Psi}_n||\mathrm{\Psi}_{n+1}|\sin(\arg\mathrm{\Psi}_n-\arg\mathrm{\Psi}_{n+1})\csc k l_{\gamma_n},\\\\
\csc kl_{\gamma_n}\equiv\displaystyle{\frac{1}{\sin kl_{\gamma_n}}}\,.
\end{array}
\end{equation}
As it can be seen from~(\ref{eq:current on the edge}) the current on edge is determined by the relation between arguments of the wave function $\Psi$ at the appropriate vertices. This situation is similar to the Josephson current~\cite{Imry97}, and, more general, to the interference of the currents in the graph. It follows from~(\ref{eq:current on the edge}) that current turns into zero when one of the absolute values is zero or if $\arg\mathrm{\Psi}_n-\arg\mathrm{\Psi}_{n+1}=\pi\,n$, $n\in \mathcal{Z}$.

Below we demonstrate the application of the proposed method on some well known quantum graphs.

\section{Scattering by the quantum ring\label{sec:Scattering by the quantum ring}}
\begin{figure}[h]
\begin{center}
\resizebox{0.25\textwidth}{!}{%
  \includegraphics{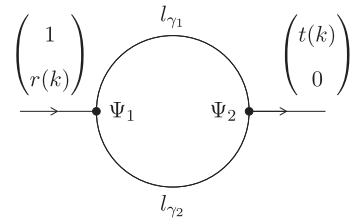}
}
\caption{The scheme of scattering by the quantum ring. The lengths of the upper and lower arms are different, namely $\lf$ and $\ls$ respectively. The form of the arm can significantly differ from the semicircle. The main demand to the curve represented the arm is the smoothness. This condition provides the existence of the tangent to this curve.}
\label{fig:quantum ring}
\end{center}
\end{figure}

In this section we consider the well known example of quantum ring~\cite{Sadreev06,Xia92,Voo06} (see eq.~(\ref{fig:quantum ring})). This model plays the same role in quantum transport as the harmonic oscillator in quantum mechanics. There are plenty of articles examined the transport properties of the quantum ring. The quantum ring is widely used (or proposed to be used) in different fields, namely spintronics~\cite{Cohen2007,Foldi2005,Xia2009}, near-field optics~\cite{Suarez2007}, optics of nanostructures~\cite{Shahbazyan2002}, Aharonov-Bohm interferometry~\cite{Hedin2005,Kokoreva09} etc.

An application of the condition~(\ref{eq:hermiticity}) at each vertex yields:
\begin{equation}\label{eq:two terminal conditions in vertex}
\begin{array}{l}
\mbox{at vertex $v_1$:}\\\\
\begin{array}{ll}
\displaystyle-[1-r(k)]+\sum_{n=1}^2&\left[\left(\GMoel{1}{11}-\GMoel{1}{21}\right)\mathrm{\Psi}_1+\right.\\\\&+\left.\left(\GMoel{1}{12}-\GMoel{1}{22}\right)\mathrm{\Psi}_2\right]=0,
\end{array}\\
\mbox{at vertex $v_2$:}\\\\
\begin{array}{l}
-t(k)+\displaystyle\sum_{n=1}^2\Big\{\left[\GMtel{1}{11}\exp(\rmi kl_{\gamma_n})-\right.\\\\ \left.-\GMtel{1}{21}\exp(-\rmi kl_{\gamma_n})\right]\mathrm{\Psi}_1+\left[\GMtel{1}{12}\exp(\rmi kl_{\gamma_n})-\right.\\\\ \left.-\GMtel{1}{22}\exp(-\rmi kl_{\gamma_n})\right]\mathrm{\Psi}_2\Big\}=0.
\end{array}
\end{array}
\end{equation}
Applying the continuity conditions~(\ref{eq:continuity}) at vertices $\mathrm{\Psi}_1$ and $\mathrm{\Psi}_2$ one obtains:
\begin{equation}\label{eq:continuity for two-terminal system}
t(k)=\mathrm{\Psi}_2, \quad 1+r(k)=\mathrm{\Psi}_1
\end{equation}
Equations~(\ref{eq:two terminal conditions in vertex}) and~(\ref{eq:continuity for two-terminal system}) represent the system of linear equations for the unknown vector of \emph{vertex amplitudes} $\mathbf{\Psi}=\left(\mathrm{\Psi}_1,\mathrm{\Psi}_2\right)^{\mathrm{T}}$
\begin{equation}\label{eq:vector equation for two-terminal system}
\mathbf{\Gamma\Psi} =\mathbf{F}, \quad \mathbf{\Gamma}=\mathbf{I}+\mathbf{\Gamma}_t,
\end{equation}
where
$$
\begin{array}{l}
(\mathbf{\Gamma}_t)_{11}=\displaystyle{\sum\limits_{n=1}^2\GMoel{n}{11}-\GMoel{n}{21}},\\\\
(\mathbf{\Gamma}_t)_{12}=\displaystyle{\sum\limits_{n=1}^2\GMoel{n}{12}-\GMoel{n}{22}},\\\\
(\mathbf{\Gamma}_t)_{21}=\displaystyle{\sum\limits_{n=1}^2\GMtel{n}{11}\exp(\rmi kl_{\gamma_n})+\GMtel{n}{21}\exp(-\rmi kl_{\gamma_n})},\\\\
(\mathbf{\Gamma}_t)_{22}=\displaystyle{\sum\limits_{n=1}^2\GMtel{n}{12}\exp(\rmi kl_{\gamma_n})+\GMtel{n}{22}\exp(-\rmi kl_{\gamma_n})},
\end{array}
$$
$$
\mathbf{I}=\left(\begin{array}{cc}1&0\\0&1\end{array}\right),\quad \mathbf{F}=\left(\begin{array}{c}2\\0\end{array}\right).
$$
The solution of~(\ref{eq:vector equation for two-terminal system}) is:
\begin{equation}
\mathbf{\Psi}=(\mathbf{I}+\mathbf{\Gamma}_t)^{-1}\mathbf{F}, \quad
\left(\begin{array}{c}
\mathrm{\Psi}_1 \\ \mathrm{\Psi}_2
\end{array}\right)=\frac{2}{\mathrm{det}\mathbf{\Gamma}}\left(\begin{array}{c}\phantom{-}\mathrm{\Gamma}_{22}\\-\mathrm{\Gamma}_{21}\end{array}\right).
\end{equation}
Finally, according to~(\ref{eq:continuity for two-terminal system}), one obtains for transmission and reflection amplitudes:
\begin{equation}
t(k)=-2\frac{\mathrm{\Gamma}_{21}}{\mathrm{det}\mathbf{\Gamma}},\quad r(k)=2\frac{\mathrm{\Gamma}_{22}}{\mathrm{det}\mathbf{\Gamma}}-1
\label{eq:amplitudes for two-terminal system}
\end{equation}
Note, that the vector $\mathbf{\Psi}$ can never be zero. The physical meaning of this fact is that simultaneous decoupling of the ring from the both external leads is not possible. This fact follows from the current conservation law.

Equation~(\ref{eq:vector equation for two-terminal system}) is the main equation of the \emph{vertex amplitudes} method. Once we construct the $\mathbf{\Gamma}$-matrix the further analysis of the problem can be completely conducted by means of its elements. The transmission and reflection amplitudes, wave function of electron, probability current density, Green's function (not presented in this paper) are all expressed through the $\mathbf{\Gamma}$-matrix elements.

The mathematical base of the method can be formulated as follows: the \emph{vertex amplitudes} method reduces the Schr\"{o}dinger equation for the quantum graph to the system of linear algebraic equation (see eq.~(\ref{eq:vector equation for two-terminal system})) for the unknown vector of vertex amplitudes $\mathbf{\Psi}=\{\mathrm{\Psi}_1,\ldots,\mathrm{\Psi}_n\}$.

\subsection{Symmetric quantum ring}
Let us begin with the most simple case corresponding to the symmetric ring (ring with equal arm lengths) with absence of any potentials on the edges. It suggests that $\mathbf{M}$ is identity matrix~\cite{Klinskikh09}:
\begin{equation}\label{eq:transfer matrix without potentials}
\mathrm{M}_{11}=\mathrm{M}_{22}=1, \mathrm{M}_{12}=\mathrm{M}_{21}=0, \zeta=\exp\left(-\rmi k l\right).
\end{equation}
Since $\mathbf{M}$ is identity matrix then $\GMo{n}\equiv\GMt{n}$. Moreover, when the edges of the ring are identical the matrices $\GMo{1}$ and $\GMo{2}$ are identical too.

Thus, the elements of $\mathbf{\Gamma}$-matrix for the ring with two equal arms of length $l$ take the following form:
\begin{equation}\label{eq:gamma matrix for the simple ring}
\begin{array}{l}
\mathbf{\Gamma}_{11}=1+2\left(\GMoel{1}{11}-\GMoel{1}{21}\right),\\\\
\mathbf{\Gamma}_{12}=2\left(\GMoel{1}{12}-\GMoel{1}{22}\right),\\\\
\mathbf{\Gamma}_{21}=2\left[\GMoel{1}{11}\exp(\rmi kl)-\GMoel{1}{21}\exp(-\rmi kl)\right],\\\\
\mathbf{\Gamma}_{22}=2\left[\GMoel{1}{12}\exp(\rmi kl)-\GMoel{1}{22}\exp(-\rmi kl)\right]-1.
\end{array}
\end{equation}
Recalling the definition of $\mathbf{\Gamma}^{(1,\gamma_n)}$ matrix~(\ref{eq:gamma matrices explicit form}) gives the explicit form of the differences in~(\ref{eq:gamma matrix for the simple ring}):
\begin{equation}\label{eq:gamma matrices differences}
\begin{array}{l}
\GMoel{1}{11}-\GMoel{1}{21}=\rmi\cot k l_{\gamma_n},\\\\
\GMoel{1}{12}-\GMoel{1}{22}=-\rmi\csc k l_{\gamma_n},\\\\
\GMoel{1}{11}\exp(\rmi kl_{\gamma_n})-\GMoel{1}{21}\exp(-\rmi kl_{\gamma_n})=\rmi\csc kl_{\gamma_n},\\\\
\GMoel{1}{12}\exp(\rmi kl_{\gamma_n})-\GMoel{1}{22}\exp(-\rmi kl_{\gamma_n})=-\rmi\cot kl_{\gamma_n}.
\end{array}
\end{equation}
Therefore, the $\mathbf{\Gamma}$-matrix for quantum ring with equal arms of length $l$ has the form:
\begin{equation}\label{eq:explicit form of gamma matrix in case of equal arms}
\mathbf{\Gamma}=\left(\begin{array}{cc}
1+2\rmi\cot kl & -2\rmi\csc kl\\
2\rmi\csc kl   & -1 - 2\rmi\cot kl
\end{array}\right).
\end{equation}
Using~(\ref{eq:amplitudes for two-terminal system}) and~(\ref{eq:explicit form of gamma matrix in case of equal arms}) we obtain the following useful for analysis expressions for the transmission and reflection amplitudes:
\begin{equation}\label{eq:transmission amplitude for the free ring}
t(k)=\frac{4\rmi}{5\sin(kl)+4\rmi\cos(kl)},
\end{equation}
\begin{equation}\label{eq:reflection amplitude for the free ring}
r(k)=-\frac{3}{5+4\rmi \cot(kl)}.
\end{equation}
As it can be seen from~(\ref{eq:transmission amplitude for the free ring}), $t(k)$ never turns into zero.

It is easy to see that values $k_n=\pi n/l$, $n=1,2,\ldots$ correspond to the zeroes of reflection amplitude.

\subsection{Asymmetric quantum ring}
The situation dramatically changes if one takes the ring with different arm lengths (asymmetric ring). In this case the $\mathbf{\Gamma}$-matrix takes form:
\begin{equation}\label{eq:gamma matrix for different edges}
\begin{array}{l}
\mathbf{\Gamma}_{11}=1+\rmi\cot k \lf+\rmi\cot k \ls,\\\\
\mathbf{\Gamma}_{12}=-\rmi\csc k\lf-\rmi\csc k\ls,\\\\
\mathbf{\Gamma}_{21}=\rmi\csc k\lf+\rmi\csc k\ls,\\\\
\mathbf{\Gamma}_{22}=-1-\rmi\cot k \lf-\rmi\cot k \ls.\\\\
\end{array}
\end{equation}
As before, the transmission and reflection amplitudes are defined by~(\ref{eq:amplitudes for two-terminal system}) and it is easy to show that:
\begin{equation}\label{eq:transmission amplitude differen edges}
\begin{array}{ll}
t(k)&=\rmi(\sin k\lf +\sin k\ls)\times\left[\frac{3}{2}\sin k\lf \sin k\ls +\right.\\\\ &+\left.\rmi\sin(k\lf+k\ls)-\cos k\lf \cos k\ls +1\right]^{-1},
\end{array}
\end{equation}
\begin{equation}\label{eq:reflection amplitude for ring with different arms}
\begin{array}{ll}
r(k)&=\Big\{\left[\sin k \lf\sin k \ls+\rmi\sin(k\lf+k\ls)\right]\times\\\\&
\times\left[\frac{3}{2}\sin k\lf\sin k\ls+\rmi\sin(k\lf+k\ls)-\right.\\\\& \left.-\cos k \lf \cos k\ls+1\right]^{-1}\Big\}-1.
\end{array}
\end{equation}

In the quantum ring transport experiments the question of resonance transmission has a fundamental aspect. By resonance transmission is meant not only the lossless passing of electron (transmission coefficient is equal to one) but also the total absence of transmission when transmission coefficient is equal to zero. The perfect transmission ($|t(k)|=1$) is observed for electron energies equal to the eigenvalues of the compact graph, in our case it is the closed ring with different arm lengths.

From~(\ref{eq:transmission amplitude differen edges}) it follows that $t(k)$ becomes zero if
\begin{equation}\label{eq:first type resonance condition}
k\ls+k\lf=2\pi n, n=1,\, 2\, \ldots,\label{eq:first type resonance condition}
\end{equation}
\begin{equation}\label{eq:second type resonance condition}
k\ls-k\lf=(2n+1)\pi, n=0,\,\pm 1,\, \pm 2 \ldots.
\end{equation}
These conditions are the same as obtained by Xia~\cite{Xia92}.

The problem of the quantum ring under consideration contains the two type of parameters: energetic parameters represented by the wave number $k$ and geometric parameters represented by the arm lengths $\lf$, $\ls$. Therefore, there are two ways to investigate the transport through the ring. In short, we could fix the arms lengths varying only the wave number and vice versa. In both cases condition~(\ref{eq:first type resonance condition}) gives the narrow (even of zero width) resonances of $t(k)$ while condition~(\ref{eq:second type resonance condition}) corresponds to the more smooth resonances with finite width. The first resonance will be referred to below as the first type resonance (\emph{FTR}), the second one -- as the second type resonance (\emph{STR}).

\begin{figure}[h]
\begin{center}
\resizebox{0.5\textwidth}{!}{%
  \includegraphics{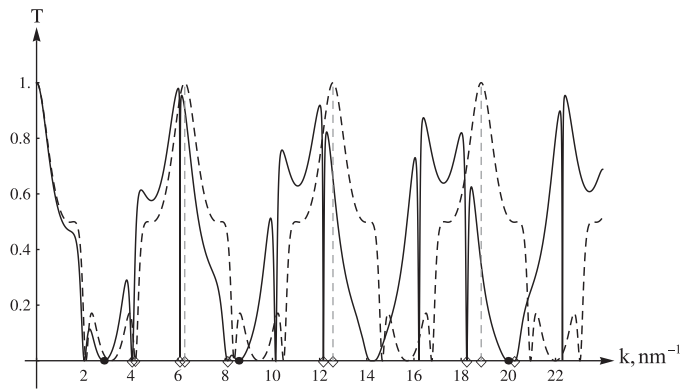}
}
\caption{Different type of the transmission resonances. The transport through the ring can be significantly different for commensurate (dashed line, $\lf=1\,\mbox{nm}$, $\ls=2\,\mbox{nm}$) and incommensurate (solid line, $\lf=1\,\mbox{nm}$, $\ls=2.1\,\mbox{nm}$) arm lengths. The diamonds correspond to the condition~(\ref{eq:first type resonance condition}), while the black points to condition~(\ref{eq:second type resonance condition}).}
\label{fig:energy resonances for ring}
\end{center}
\end{figure}

Let us first consider the energy dependence of the transmission amplitude (see fig.~\ref{fig:energy resonances for ring}).
Transmission coefficient demonstrates characteristic peculiarities in the case of incommensurate lengths for the values of $k$ satisfying the condition~(\ref{eq:first type resonance condition}). Namely, the very narrow dips. Such behavior can be explained by analysis of the series expansion of the transmission amplitude $t(k)$. For instance, near the point $\displaystyle{k=\frac{6\pi}{3.1}}$, marked in fig.~\ref{fig:energy resonances for ring} by the diamond, it has the following form:
\begin{equation}\label{eq:resonance form}
t(k)\approx \frac{\beta \Delta k}{\Delta k + \mathrm{\Omega}}, \quad \Delta k=k-\displaystyle{\frac{6\pi}{\lf+\ls}}\approx k-6.08,
\end{equation}
where
$$
\beta=2\rmi\displaystyle\frac{\left[-1+2\cos\left(2\pi\displaystyle\frac{\ls-\lf}{\ls+\lf}\right)\right]
\cos\left(\pi\displaystyle\frac{\ls-\lf}{\ls+\lf}\right)}{4-\rmi\displaystyle\frac{\ls-\lf}{\ls+\lf}\sin\left(6\pi\displaystyle\frac{\ls-\lf}{\ls+\lf}\right)},
$$
and
$$
\mathrm{\Omega}=-\displaystyle\rmi\frac{-1+\cos\left(6\pi\displaystyle\frac{\ls-\lf}{\ls+\lf}\right)}{4(\ls+\lf)-\rmi(\ls-\lf)\sin\left(6\pi\displaystyle\frac{\ls-\lf}{\ls+\lf}\right)}.
$$

\begin{figure}[h]
\begin{center}
\resizebox{0.48\textwidth}{!}{%
  \includegraphics{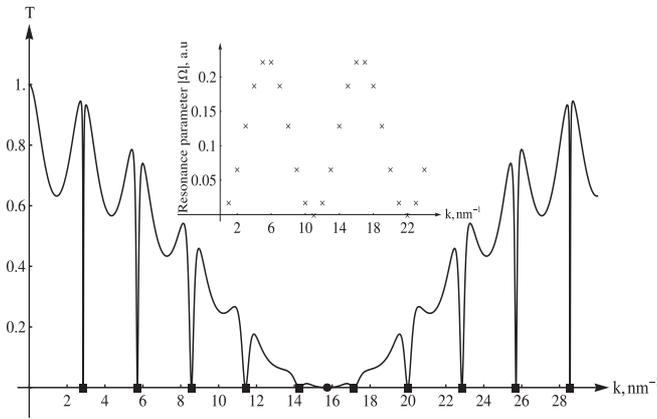}
}
\caption{The dependence of the \emph{FTR} width on the electron wave number $k$, $\ls=1\,\mbox{nm}$, $\lf=1.1\,\mbox{nm}$. The black squares show the position of the \emph{FTR}, black points correspond to the \emph{STR}. The periodical variation of the parameter $|\mathrm{\Omega}|$ in~(\ref{eq:resonance form}) is illustrated in the inset.}
\label{fig:resonance width}
\end{center}
\end{figure}

The parameter $\mathrm{\Omega}$ in the denominator of this expansion determine the width of the resonance. One can easily obtain that for commensurate lengths ($\ls=1\,\mbox{nm}$, $\lf=2\,\mbox{nm}$) this term is equal to zero and amplitude $t(k)$ is equal to one. In opposite case of the incommensurate lengths (namely,$\ls=1\,\mbox{nm}$, $\lf=2.1\,\mbox{nm}$) the parameter $\mathrm{\Omega}$ is nonzero and in fig.~\ref{fig:energy resonances for ring} we see the narrow resonance. Figure~\ref{fig:energy resonances for ring} illustrates that the resonances width grows with increasing of $k$. In fact, the resonance width changes periodically upon varying the $k$ (see fig.~\ref{fig:resonance width}). The reduction of the transmission coefficient, similar to presented in fig.~\ref{fig:resonance width}, was revealed by Wu and Mahler~\cite{Wu91} in studies of Aharonov-Bohm effect in metals and semiconductors.
The \emph{STR} position are marked by the black points in fig.~\ref{fig:energy resonances for ring} and~\ref{fig:resonance width}. One can see that they are significantly smoother compared with the \emph{FTR}. There width is always finite for a finite values of $k$.

\begin{figure}[h]
\begin{center}
\resizebox{0.48\textwidth}{!}{%
  \includegraphics{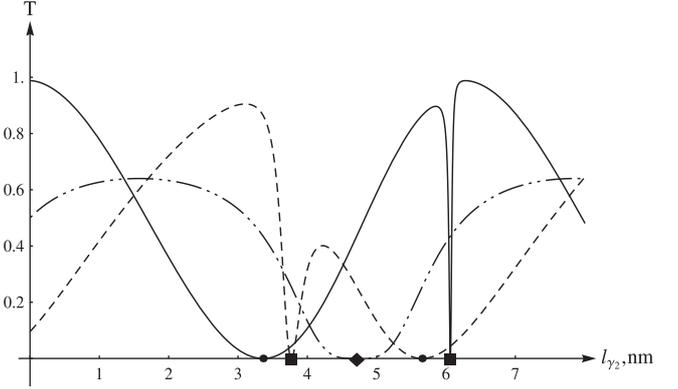}
}
\caption{The transmission coefficient versus second arm length. The general view of the transmission coefficient resembles the one presented in Fig.~\ref{fig:energy resonances for ring}. One can see the characteristic form of the \emph{FTR}. Their positions are marked by the squares. Depending on the upper arm length the width of these resonances is changing, $\lf=0.22\,\mbox{nm}$ (solid line), $\lf=0.8\pi\,\mbox{nm}$ (dashed line). Dashed-dotted  line ($\lf=\pi/2\,\mbox{nm}$) presents the situation when the lengths of the arms are equal. When this condition takes place the transmission coefficient turns into zero (black diamond in the $x$-axis). The electron wave number is fixed to be $k=1\,\mbox{nm}^{-1}$.}
\label{fig:resonances for ring}
\end{center}
\end{figure}

Now, let us consider the ``geometric'' resonances conditioned by the dependence of the transmission amplitude on the ring lengths. The wave number of electron is fixed in this case. As it was noted above there are two different type of resonances (see fig.~\ref{fig:resonances for ring}). The forms of resonances resemble the ones considered above (see fig.~\ref{fig:energy resonances for ring}). Nevertheless, there is important distinction between these two cases. Expanding the numerator and denominator
of $t(k)$ defined in~(\ref{eq:transmission amplitude differen edges}) in powers of $k \ls$ up to the first-order terms near two points $k\ls=2\pi-k\lf$ and $k\ls=\pi+k\lf$ we obtain:
\begin{equation}\label{eq:series expansion of transmission amplitude}
\begin{array}{l}
t(k)\approx \displaystyle\frac{\rmi\Delta y\cos x}{\Delta y \left(\frac{1}{4} \sin 2x +\rmi\right)-\frac{1}{2}\sin^2 x},\, \Delta y=y+x-2\pi;\\\\
t(k)\approx\\\\ \displaystyle\frac{\rmi\Delta y\cos x}{\Delta y \left(\frac{5}{4} \sin 2x - \rmi\cos 2x\right)+\left(\frac{3}{2}\cos ^2 x-\frac{1}{2}\cos 2x - \rmi \sin2 x\right)},\\\\
\Delta y=y-x+\pi,
\end{array}
\end{equation}
where $x\equiv k\lf$, $y\equiv k\ls$. The main difference of two expansions~(\ref{eq:series expansion of transmission amplitude}) consists in behavior at the vicinity of the points $x=\pi n$, $n=1,2,\ldots$. It can be obtained directly from~(\ref{eq:series expansion of transmission amplitude}) that the limits $\Delta y\to 0$ and $x\to \pi n$ do not commutate in the first case:
\begin{equation}\label{eq:noncommutative limits for amplitude}
\begin{array}{c}
\displaystyle\lim_{x\to\pi n}\lim_{\Delta y\to 0} \left\{t(k),r(k)\right\}=\left\{0,1\right\},\\\\
\displaystyle\lim_{\Delta y\to 0} \lim_{x\to\pi n} \left\{t(k),r(k)\right\}=\left\{(-1)^n,0\right\},
\end{array}
\end{equation}
and commutate in the second. This non-commutativity is even more sharply manifested for coefficients of the wave function $(a_1,\,b_1)$, $(a_2,\,b_2)$
\begin{equation}\label{eq:noncommutative limits for coefficients}
\begin{array}{c}
\displaystyle\lim\limits_{x\to\pi n}\lim\limits_{\Delta y\to 0} a_1=\mathrm{const},\\\\
\displaystyle\lim\limits_{\Delta y\to 0} \lim\limits_{x\to\pi n} a_1=\rmi\infty.
\end{array}
\end{equation}
Bulgakov~\textit{et.al.}~\cite{Sadreev06} associate such a behavior of the transmission amplitude and the wave function in the ring with the existence of the bounded states in continuum. In details it is discussed below.

The noncommutative properties of the wave function coefficients and transmission amplitude can be understood using the $\mathbf{\Gamma}$-matrix~(\ref{eq:gamma matrix for different edges}). On the one hand, if $k\ls=2\pi n-k\lf$ then $\mathbf{\Gamma}_{t}\equiv 0$, on the other hand $k\lf=\pi n$ are the points of singularity of all elements of $\mathbf{\Gamma}_{t}$. This fact shed light to non-commutativity of the two aforementioned limits~(\ref{eq:noncommutative limits for coefficients}).

While $k\ls=2\pi n-k\lf$ make the $\mathbf{\Gamma}_{t}$ be zero matrix, the values $k\ls=\pi n+k\lf$ keep diagonal elements nonzero. The diagonal elements in this case, $\pm2\rmi \cot k\lf$, determine the phase of the reflection amplitude, namely $\varphi\equiv\arg[r(k)+1]$:
\begin{equation}\label{eq:reflection argument for resonances}
\tan \varphi=-2\cot k\lf.
\end{equation}
It can be useful to note that at the same time $\varphi$ is the phase of the wave function at vertex $v1$ according to~(\ref{eq:continuity for two-terminal system}).

\subsection{Asymmetric Aharonov-Bohm ring}
The problem becomes more complicated if at the center of the ring there is an infinitely thin solenoid carrying finite magnetic flux. At the same time it is a standard situation in the quantum transport~\cite{Sadreev06}.

\begin{figure}[h]
\begin{center}
\resizebox{0.3\textwidth}{!}{%
  \includegraphics{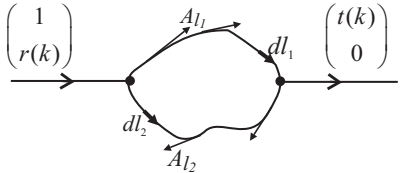}
}
\caption{The quantum loop with arms of arbitrary form coupled to the two leads. The loop is penetrated by the Aharonov-Bohm flux. One can see that directions of tangential component of the vector potential $\mathbf{A}$ and vector $\mathbf{dl}_{\gamma_n}$ are opposite on the upper and lower arms.}
\label{fig:loop with flux inside}
\end{center}
\end{figure}

Let us consider the asymmetric quantum ring with different arms of length $\lf$ and $\ls$ penetrated by the magnetic flux $\mathrm{\Phi}$. From the mathematical point of view it is picewise-smooth contour. Thus, it can be parameterized by the two natural parameters, namely, $\xi_{\gamma_1}$ and $\xi_{\gamma_2}$. We choose the ``solenoid'' in such a way that the vector potential $\mathbf A$ has only tangent to the contour component (see fig.~\ref{fig:loop with flux inside}).

The Stokes' theorem gives
$$
\oint\limits_L \mathbf{A}\mathbf{dl} =\mathrm{\Phi}.
$$
According to our choice of the vector potential
$$
\oint\limits_L \mathbf{A}\mathbf{dl} =A_{\lf}\lf+A_{\ls}\ls .
$$
If one takes $A_{\lf}=A_{\ls}=A$, then
$$
A = \displaystyle{\frac{\mathrm{\Phi}}{\ls+\lf}}.
$$

Thus, the magnetic flux modifies the phase of the wave function in the ring while the \textit{in,out}-wave functions are still the same:
\begin{equation}\label{eq:wave function under aharonov-bohm}
\begin{array}{l}
\psi_{in}=\exp(\rmi k x)+r(k)\exp(-\rmi k x),\\\\
\psi_{out}=t(k)\exp(\rmi k x),\\\\
\psin{1} =a_{\gamma_1}\exp(\rmi k^- x)+b_{\gamma_1}\exp(-\rmi k^+ x),\\\\
\psin{2} =a_{\gamma_2}\exp(\rmi k^+ x)+b_{\gamma_2}\exp(-\rmi k^- x),
\end{array}
\end{equation}
where $k^-=k-\alpha$, $k^+=k+\alpha$, $\alpha=\mathrm{\Phi}/(\mathrm{\Phi}_0 L)$, $\mathrm{\Phi}$ is the magnetic flux through the loop section area, $\mathrm{\Phi}_0=2\pi\hbar c/e$ and $L=\lf+\ls$.

Applying the developed in previous sections method, one obtains the following expressions for the elements of matrices $\GMo{1}$, $\GMo{2}$
\begin{equation}\label{eq:gamma matrices in case of Aharonov-Bohm}
\begin{array}{l}
\GMo{1}_{11}=\displaystyle\frac{\exp(-\rmi k^+ l_{\gamma_1})}{\exp(-\rmi k^+ l_{\gamma_1})-\exp(\rmi k^- l_{\gamma_1})},\\\\
\GMo{1}_{12}=\displaystyle\frac{-1}{\exp(-\rmi k^+ l_{\gamma_1})-\exp(\rmi k^- l_{\gamma_1})},\\\\
\GMo{1}_{21}=\displaystyle\frac{-\exp(\rmi k^- l_{\gamma_1})}{\exp(-\rmi k^+ l_{\gamma_1})-\exp(\rmi k^- l_{\gamma_1})},\\\\
\GMo{1}_{22}=\displaystyle\frac{1}{\exp(-\rmi k^+ l_{\gamma_1})-\exp(\rmi k^- l_{\gamma_1})},\\\\
\GMo{2}_{11}=\displaystyle\frac{\exp(-\rmi k^- l_{\gamma_2})}{\exp(-\rmi k^- l_{\gamma_2})-\exp(\rmi k^+ l_{\gamma_2})},\\\\
\GMo{2}_{12}=\displaystyle\frac{-1}{\exp(-\rmi k^- l_{\gamma_2})-\exp(\rmi k^+ l_{\gamma_2})},\\\\
\GMo{2}_{21}=\displaystyle\frac{-\exp(\rmi k^+ l_{\gamma_2})}{\exp(-\rmi k^- l_{\gamma_2})-\exp(\rmi k^+ l_{\gamma_2})},\\\\
\GMo{2}_{22}=\displaystyle\frac{1}{\exp(-\rmi k^- l_{\gamma_2})-\exp(\rmi k^+ l_{\gamma_2})},
\end{array}
\end{equation}
and $\mathbf{\Gamma}$
\begin{equation}\label{eq:gamma matrix in case of Aharonov-Bohm}
\begin{array}{l}
\mathbf{\Gamma}_{11}=1+\rmi\cot k\lf+\rmi\cot k\ls,\\\\
\mathbf{\Gamma}_{12}=-\rmi\exp(\rmi \alpha \lf)\csc k\lf -\rmi\exp(-\rmi \alpha \ls)\csc k\ls,\\\\
\mathbf{\Gamma}_{21}=\rmi\exp(-\rmi \alpha \lf)\csc k\lf +\rmi\exp(\rmi \alpha \ls)\csc k\ls,\\\\
\mathbf{\Gamma}_{22}=-1-\rmi\cot k\lf-\rmi\cot k\ls.
\end{array}
\end{equation}

Note that the same matrix was used by Texier and B\"{u}ttiker in studies of the quantum ring with the Aharonov-Bohm flux~\cite{Texier2003}.

From~(\ref{eq:amplitudes for two-terminal system}),~(\ref{eq:gamma matrix in case of Aharonov-Bohm}) we obtain the transmission amplitude for the ring with different arms $\lf$, $\ls$ penetrated by the magnetic flux $\Phi$:
\begin{equation}\label{eq:transmission amplitude in case of Aharonov-Bohm}
t(k)=2\rmi\frac{\exp(\rmi\alpha\lf)\sin k\ls+\exp(-\rmi\alpha\ls)\sin k\lf}{\sin k\lf\sin k\ls \det\mathbf{\Gamma}}.
\end{equation}
Comparing this with~(\ref{eq:transmission amplitude differen edges}), we understand that the presence of the new parameter $\alpha$ modifies the conditions of the transmission amplitude zeros~(\ref{eq:first type resonance condition}). One can see from~(\ref{eq:gamma matrix in case of Aharonov-Bohm}) that $\alpha$ keeps the same points of $\mathbf{\Gamma}$-matrix singularity but changes condition~(\ref{eq:first type resonance condition}). Straightforward calculation shows that if $k\ls+k\lf=2\pi$ then
\begin{equation}\label{}
\Gamma_{21}=\csc k\lf \exp\left(-\rmi\alpha\ls\right)\left[1-\exp\left(\rmi\displaystyle{\frac{\mathrm{\Phi}}{\mathrm{\Phi}_0}}\right)\right].
\end{equation}
The condition $\mathrm{\Gamma}_{21}=0$ gives $\mathrm{\Phi}/\mathrm{\Phi}_0=2\pi n$ or $\alpha L=2\pi n$,  where $n=0,\pm1,\pm2,\ldots$. Thus, to observe the \emph{FTR} the two simultaneous conditions is needed to take place:
\begin{enumerate}
\item $k\ls+k\lf=2\pi n$, $n= 1, 2, \ldots\;$;
\item $\alpha L=2\pi m$, $\;m=0,\pm 1, \pm 2, \ldots\;$.
\end{enumerate}
It can be directly verified that in the case of symmetric ring these conditions and ones obtained by Bulgakov~\textit{et.al.}~\cite{Sadreev06} appears to be the same.

\begin{figure}[h]
\begin{center}
\resizebox{0.45\textwidth}{!}{%
  \includegraphics{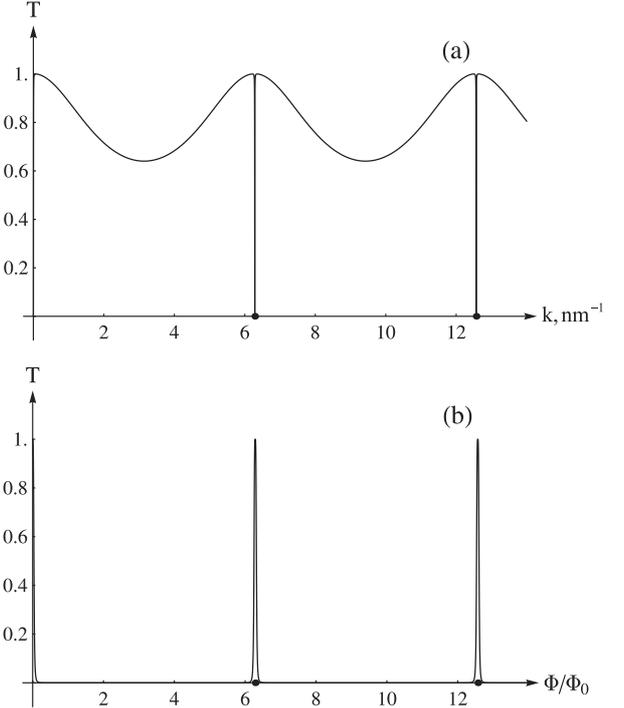}
}
\caption{The nonuniform nature of the transmission amplitude in the vicinity of the \emph{FTR}. The arms are of the same length, $\lf=\ls=1/2\,\mbox{nm}$. The conditions of the \emph{FTR} are formulated above and coincide with the ones obtained by Bulgakov~\textit{et.al.}~\cite{Sadreev06}. It is seen that the limits $\Delta k\to 0$ and $\Delta \alpha \to 0$ do not commutate: (a) $\lim\limits_{\Delta k\to 2\pi}\lim\limits_{\Delta \alpha \to 2\pi}=0$, (b) $\lim\limits_{\Delta \alpha\to 2\pi}\lim\limits_{\Delta k\to 2\pi}=1$. Here $\Delta k=k-k_0$, $\Delta \alpha=\alpha-\alpha_0$ and $k_0,\,\alpha_0$ are the points of \emph{FTR}. For example, these can be $k_0=2\pi$, $\alpha_0=2\pi$.}
\label{fig:noncommutativity}
\end{center}
\end{figure}

The noncommutative properties of the transmission amplitude emerge upon calculation the limits $\Delta k\to 0$ and $\Delta \alpha \to 0$. For the ring with equal arms it was done by Bulgakov~\textit{et.al.}~\cite{Sadreev06}. Figure~\ref{fig:noncommutativity} illustrates this noncommutative behavior of transmission amplitude.

\begin{figure}[h]
\begin{center}
\resizebox{0.5\textwidth}{!}{%
  \includegraphics{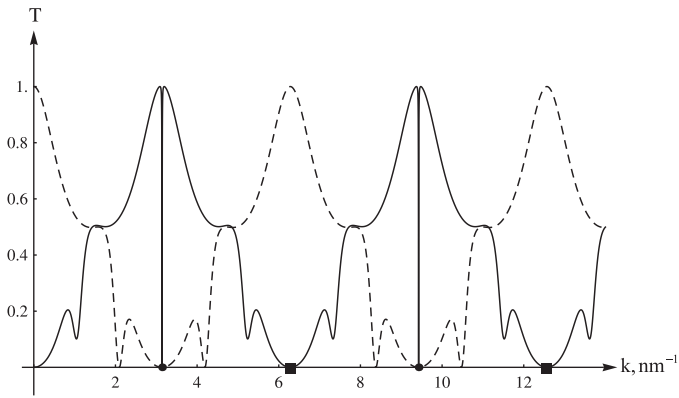}
}
\caption{The illustration of the induced by the Aharonov-Bohm flux change of the resonance type and appearance of new resonances. The black squares correspond to the condition~(\ref{eq:first type resonance condition}), the black points correspond to the condition~(\ref{eq:second type resonance condition}). Dashed line presents the transmission coefficient for $\lf=1\,\mbox{nm}$, $\ls=2\,\mbox{nm}$, $\alpha=0$, solid line -- $\lf=1\,\mbox{nm}$, $\ls=2\,\mbox{nm}$, $\alpha=3$. Thereby, the Aharonov-Bohm conditions induce the \emph{FTR} even for commensurate arms lengths.}
\label{fig:resonance type changes}
\end{center}
\end{figure}

One feature needs to be recognized. When one takes $\alpha L=2\pi m$, $\;m=0,\pm 1, \pm 2, \ldots$  the value of the transmission amplitude differs from ones with $\alpha L=0$, but transmission coefficient is the same. From the other hand, the values of the parameter $\alpha\neq 2\pi m$ can lead to the change of the resonance type or appearance of new resonances (see fig.~\ref{fig:resonance type changes}).

\begin{figure}[h]
\begin{center}
\resizebox{0.48\textwidth}{!}{%
  \includegraphics{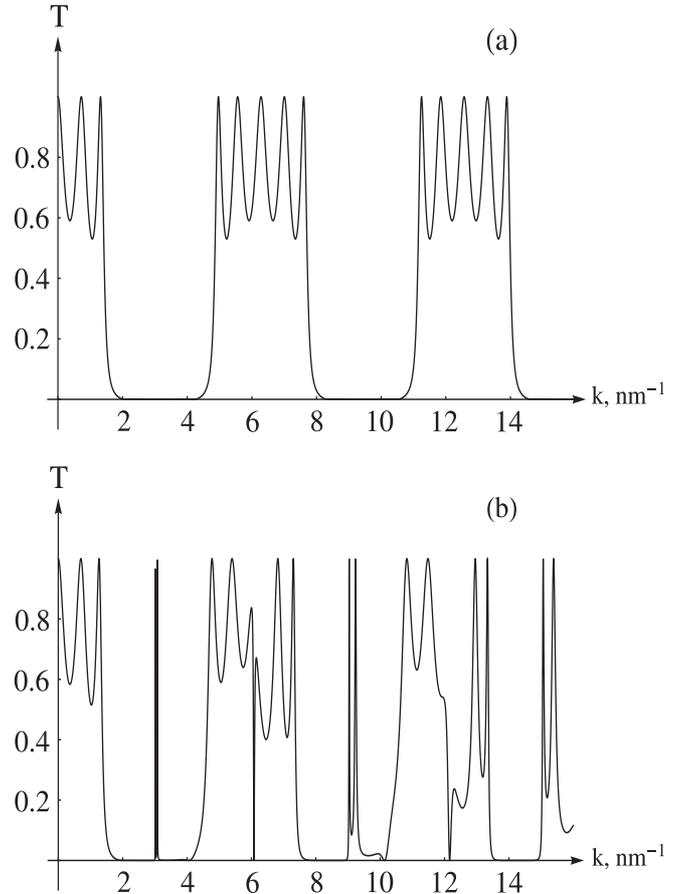}
}
\caption{This figure shows the transmission coefficient through the linear chain of the connected quantum rings. The line-shape can be rather different depending on the ratio of the arm lengths: (a) transmission coefficient for the three quantum ring with different, but commensurate arm lengths $\lf=1\,\mbox{nm}$, $\ls=2\,\mbox{nm}$; (b) transmission coefficient for the three quantum ring with different and incommensurate arm lengths $\lf=1\,\mbox{nm}$, $\ls=2.1\,\mbox{nm}$.}
\label{fig:quantum rings chain}
\end{center}
\end{figure}

\subsection{Scattering by the cascade of quantum rings}
Let us consider the quantum transport in the cascade of quantum rings. It is well known that convenient method to work with one-dimensional quantum cascades is the transfer matrix method~\cite{Merz70,Klinskikh08}. Transfer matrix of the linear chain of $N$ quantum rings and transmission amplitude of this cascade are calculated as:
\begin{equation}
\begin{array}{c}
\mathbf{M}^{tot}=\mathbf{M_1}\ldots\mathbf{M_N},\\\\
t(k)=\left[\mathrm{M}^{tot}_{11}\right]^{-1},
\end{array}
\end{equation}
where $\mathbf{M_1},\,\ldots\mathbf,\,\mathbf{M_N}$ are the transfer matrices of intermediate rings, $\mathbf{M}^{tot}$ is the transfer matrix of the chain and $t(k)$ is the transmission amplitude of the chain.

Using expressions for transmission and reflection amplitudes~(\ref{eq:transmission amplitude for the free ring}),~(\ref{eq:reflection amplitude for the free ring}),~(\ref{eq:transmission amplitude differen edges}),~(\ref{eq:reflection amplitude for ring with different arms}),~(\ref{eq:transmission amplitude in case of Aharonov-Bohm}) one can construct the transfer matrix of appropriate quantum ring (symmetric or asymmetric) according to the definition~(\ref{eq:general form of transfer matrix}).

In fig.~\ref{fig:quantum rings chain} the transmission coefficient for the different chains of quantum rings as a function of the electron wave number is presented.

\section{Conclusion\label{sec:Conclusions}}
In this paper, we have demonstrated the successive analytical approach to the solution of the transport problem for a quantum graph. Within the framework of the developed approach the solution of the transport problem is equivalent to the solution of a linear system of equations for the \emph{vertex amplitudes} $\mathbf{\Psi}$. All major properties, such as transmission and reflection amplitudes, wave function on the graph, probability current, then are expressed in terms of one $\mathbf{\Gamma}$-matrix that determines the transport through the graph.

The simplicity and convenience of the proposed method have been illustrated on the example of quantum ring on the different assumptions. The compact part of the graph under consideration has the two-terminal structure. However, the possibility of the expansion of the \emph{vertex amplitudes} method on the many-terminal systems is obvious. This will resulted only in the increasing of the size of the $\mathbf{\Gamma}$-matrix .

The carried out analysis revealed the two kind of resonances upon transport through the ring. The most interesting is the first one. The nonuniform behavior of the transmission amplitude (see~\ref{eq:noncommutative limits for amplitude}), features of the resonance width, the specific structure of the resonance in the complex energy plane (not presented here) make the \emph{FTR} very similar to the BIC considered by Pursey and Webber~\cite{Pursey1995}. In this paper we intentionally avoided the use of term BIC for the quantum ring. From our point of view, the direct analogy with the results of von Neuman and Wigner~\cite{Wigner1929}, Pursey and Webber~\cite{Pursey1995} is not correct in the case of the quantum ring without any potentials. One of the central idea of von Neuman and Wigner was in special choice of the potential the electron is subjected to. This choice leads to the integrability of the continuum wave function. The relationship between the properties of the  wave function and potential is crucial in the definition of the BIC. For the quantum ring we observe the emergence of the \emph{FTR} (in both cases of rings with and without Aharonov-Bohm flux) in the absence of any potential.

The transmission resonances for the quantum graphs and, in particular, for the quantum ring, were studied in a number of works~\cite{Hedin2005,Shao1994,Satanin2005,Deo2006}. Upon studying the resonance line-shape features it was established these resonances are the Fano resonances. The Fano resonances appear when the coupling between the discrete spectrum and continuum takes place. Most often, the discrete spectrum originates from the transverse quantization. However, even in a more simple quasi-one-dimensional model the Fano resonances emerge~\cite{Shao1994}. In this case, they come from the coupling of quasi-bound states of the compact part of the graph and continuum of the external leads. This in turn gives rise to the peculiarities of the the resonance structure in the complex energy plane~\cite{Shao1994}. The classical analog of this phenomena are the Helmholtz resonances~\cite{Rossing2007}. The ring play the role of the resonator connected to the waveguides. The quasi-bound states can leak out from the ring, that is the coupling is done by the contacts.

Let us notice that the theoretical framework developing during research of quantum graph is interesting for a field that are, at first glance, far from the considered problem. These are neurobiology and cellular biology. The first reason follows from the usability of the graph to establish the relationships between the different parts of complex system. The modular constitution of biological systems becoming more and more convenient~\cite{Hartwell99}. Thus, the action potentials characteristic of nerve and muscle cells have been reconstituted by transplanting ion channels and pumps from such cells into non-excitable cells~\cite{Hsu93}. Thereby, the properties of single modules and molecular connections between them are similar to electrical circuits (with the Kirchhoff rules) and quantum graphs (with the principles of quantum mechanics).

The second reason concerned with the existence of biological systems that are the natural graphs. For example, it is neural networks in brain~\cite{Sporns2004,Iturria2008,Galarreta99}. The fundamental properties of neural networks strongly depend on its topology and coupling conditions at the nodes. Note, that the same is typical for the quantum graphs.
\appendix

\section{\label{app:potential}Scattering by the potential with finite support}
\begin{figure}[h]
\begin{center}
\resizebox{0.21\textwidth}{!}{%
  \includegraphics{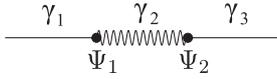}
}
\caption{Schematic representation of the scattering by the potential with finite support. Wavy line represents the potential.}
\label{fig:scattering by arbitrary potential}
\end{center}
\end{figure}

By way of illustration of the \emph{vertex amplitudes} method, let us consider a well known problem of electron scattering on the one-dimensional potential with finite support (see fig.~\ref{fig:scattering by arbitrary potential}). Below, we derive some general expressions for transmission and reflection amplitudes using the \emph{vertex amplitudes} approach. It must be pointed out that the obtained expressions are valid for the potentials with arbitrary profiles (not only for the square barriers, wells etc.).
Let us consider the left scattering problem in the infinite quasi-one-dimensional lead with real potential $V(x)\in C_0^{\infty}$. In fact, these requirements are more stringent than those actually needed for the validity of the results reported below. Following the graph ideology we treat this system as a graph consisting of three edges $\gamma_1$, $\gamma_2$, $\gamma_3$ and two vertices $v_1$, $v_2$. The local coordinates $\xin{1}\in(-\infty,\xi^{(1)}_{v_1}]$, $\xin{2}\in[\xi^{(2)}_{v_1},\xi^{(2)}_{v_2}]$ and $\xin{3}\in[\xi^{(3)}_{v_2},+\infty )$ are defined in their respective edges. Let potential $V(x)$ to be located on the edge $\gamma_2$. For simplicity, we assume that the electron's effective mass $m_{\gamma_n}$ to be equal to the mass of the free electron. According to the scattering theory ideology the wave functions in the edges $\gamma_{1}$, $\gamma_3$ are taken in the form~(\ref{eq:psi-in,out for left scattering}).

According to theory of differential equations~\cite{Coddington55}, there are two linearly independent solutions $u(x)$ and $v(x)$ of the Schr\"{o}dinger equation in the edge $\gamma_2$. Thus, the wave function $\psin{2}$ can be chosen as a linear combination of functions $u$ and $v$, namely:
\begin{equation}
\psin{2}(\xin{2})=a\,u(\xin{2})+b\,v(\xin{2}).
\end{equation}
Let us express the coefficients $(a,b)$ of the $\psin{2}$ in terms of $\Psi_1$ and $\Psi_2$, where $\Psi_1$ and $\Psi_2$ are the values of the electron wave function at the vertices $v_1$ and $v_2$ respectively. To do this, we use the obvious relations:
\begin{equation}
\begin{array}{l}
a\,u(\xi^{(2)}_{v_1}) + b\,v(\xi^{(2)}_{v_1})=\mathrm{\Psi}_1,\\\\
a\,u(\xi^{(2)}_{v_2}) + b\,v(\xi^{(2)}_{v_2})=\mathrm{\Psi}_2.
\end{array}
\end{equation}
Solving this system one obtains:
\begin{equation}\label{eq:alpha beta vs. psi1 psi2}
\begin{array}{l}
\left(\begin{array}{c}a\\b\end{array}\right)=\displaystyle{\frac{1}{u(\xi^{(2)}_{v_1})v(\xi^{(2)}_{v_2})-
u(\xi^{(2)}_{v_2})v(\xi^{(2)}_{v_1})}}\times\\\\
\times\displaystyle{\left(\begin{array}{cc}\phantom{-}v(\xi^{(2)}_{v_2})&-v(\xi^{(2)}_{v_1})\\
-u(\xi^{(2)}_{v_2})&\phantom{-}u(\xi^{(2)}_{v_1})\end{array}\right)\left(\begin{array}{c}\mathrm{\Psi}_1\\\mathrm{\Psi}_2\end{array}\right)}.
\end{array}
\end{equation}
Applying the hermiticity conditions~(\ref{eq:hermiticity}) to the wave function at vertices $v_1$, $v_2$ one obtain the two equations:
\begin{equation}\label{eq:equations for trivial graph1}
\rmi k\exp(\rmi k \xivn{1}^{(1)})[1-r(k)]-\left[a u'(\xivn{1}^{(2)})+b v'(\xivn{1}^{(2)})\right]=0,
\end{equation}
\begin{equation}\label{eq:equations for trivial graph2}
-\rmi k\, t(k)\exp(\rmi k \xivn{2}^{(3)})+\left[a u'(\xivn{2}^{(2)})+b v'(\xivn{2}^{(2)})\right]=0,
\end{equation}
where a prime denotes differentiation with respect to appropriate local coordinate. Using the continuity conditions~(\ref{eq:continuity}) in vertices and~(\ref{eq:alpha beta vs. psi1 psi2}) we obtain the system of two linear equations for unknown vector of \emph{vertex amplitudes} $\mathbf{\Psi}=\left(\mathrm{\Psi}_1,\mathrm{\Psi}_2\right)^{\mathrm{T}}$ that can be written in the matrix form:
\begin{equation}\label{eq:system for arbitrary potential}
\mathbf{\Gamma\Psi}=\mathbf{F},
\end{equation}
where
$$
\begin{array}{c}
\begin{array}{l}
\mathbf{\Gamma}_{11}=u'(\xivn{1}^{(2)})v(\xivn{2}^{(2)})-u(\xivn{2}^{(2)})v'(\xivn{1}^{(2)}) + \rmi k \widetilde{\mathrm{\Delta}},\\\\
\mathbf{\Gamma}_{12}=u(\xivn{1}^{(2)})v'(\xivn{1}^{(2)})-v(\xivn{1}^{(2)})u'(\xivn{1}^{(2)}),\\\\
\mathbf{\Gamma}_{21}=v'(\xivn{2}^{(2)})u(\xivn{2}^{(2)})-u'(\xivn{2}^{(2)})v(\xivn{2}^{(2)}),\\\\
\mathbf{\Gamma}_{22}=u'(\xivn{2}^{(2)})v(\xivn{1}^{(2)}) -u(\xivn{1}^{(2)})v'(\xivn{2}^{(2)}) + \rmi k \widetilde{\mathrm{\Delta}},
\end{array}\\\\
\mathbf{F}=\left(\begin{array}{c}2\rmi k \widetilde{\mathrm{\Delta}}\exp(\rmi k \xivn{1}^{(1)})\\0\end{array}\right),\\\\
\widetilde{\mathrm{\Delta}}=u(\xivn{1}^{(2)})v(\xivn{2}^{(2)})-u(\xivn{2}^{(2)})v(\xivn{1}^{(2)}).
\end{array}
$$
The solution of~(\ref{eq:system for arbitrary potential}) is:
\begin{equation}\label{eq:solution for arbitrary potential}
\begin{array}{c}
\mathbf{\Psi}=(\rmi k\widetilde{\mathrm{\Delta}}\mathbf{I}+\mathbf{\Gamma}_t)^{-1}\mathbf{F},\\\\
\displaystyle\left(\begin{array}{c}
\mathrm{\Psi}_1 \\ \mathrm{\Psi}_2
\end{array}\right)=\frac{2\rmi k\exp(\rmi k \xivn{1}^{(1)})\widetilde{\mathrm{\Delta}}}{\mathrm{det}\mathbf{\Gamma}}\left(\begin{array}{c}\phantom{-}\mathrm{\Gamma}_{22}\\-\mathrm{\Gamma}_{21}\end{array}\right).
\end{array}
\end{equation}
The transmission and reflection amplitudes are easily calculated according to~(\ref{eq:conditions at in out vertices}):
\begin{equation}\label{eq:continiuty conditions for arbitrary potential}
\begin{array}{l}
t(k)=\mathrm{\Psi}_2\exp(-\rmi k \xivn{2}^{(3)}),\\\\
r(k)=\Big[\mathrm{\Psi}_1-\exp(\rmi k \xivn{1}^{(1)})\Big]\exp(\rmi k \xivn{1}^{(1)}).
\end{array}
\end{equation}
Substitution of~(\ref{eq:solution for arbitrary potential}) into~(\ref{eq:continiuty conditions for arbitrary potential}) gives:
\begin{equation}\label{eq:transmission amplitude for arbitrary potential}
t(k)=\displaystyle{\frac{2\rmi\,k\left(\nu_2-\mu_2\right)\exp[-\rmi k (\xivn{2}^{(3)}-\xivn{1}^{(1)})]}{\displaystyle{\frac{u_1}{u_2}(\nu_2-\rmi k)(\mu_1+\rmi k) -  \frac{v_1}{v_2}(\nu_1-\rmi k)(\mu_2+\rmi k)} } },
\end{equation}
where
$$
\begin{array}{c}
\left(u_{1,2};v_{1,2}\right)\equiv\left(u(\xivn{1,2}^{(2)});v(\xivn{1,2}^{(2)})\right),\\\\
\left(\mu_{1,2};\nu_{1,2}\right)\equiv\left(\displaystyle\frac{u'(\xi)}{u(\xi)}\Bigg|_{\xi=\xivn{1,2}^{(2)}};\frac{v'(\xi)}{v(\xi)}\Bigg|_{\xi=\xivn{1,2}^{(2)}}\right).
\end{array}
$$

The expression for the reflection amplitude can be obtained in analogous manner, namely:
\begin{equation}\label{eq:reflection amplitude for arbitrary potential}
\begin{array}{ll}
r(k)&=\displaystyle\frac{u_2 v_1(\mu_2-\rmi k)(\nu_1-\rmi k)-u_1v_2(\nu_2-\rmi k)(\mu_1-\rmi k)}{u_1 v_2(\nu_2-\rmi k)(\mu_1+\rmi k)-u_2v_1(\nu_1+\rmi k)(\mu_2-\rmi k)}\times\\\\&\times\exp\left(2\rmi k \xivn{1}^{(1)}\right).
\end{array}
\end{equation}
The transfer-matrix of potential $V(x)$ could be written down according to~(\ref{eq:general form of transfer matrix}).

It is significant to point out that the $\mathbf{\Gamma}$-matrix carries information about discrete spectrum of potential $V(x)$. Using the standard method~\cite{LandauIII}, or directly from the properties of the transmission amplitude~\cite{Novikov84} it follows:
\begin{equation}\label{eq:discrete spectra and gamma matrix}
\frac{\det\mathbf{\Gamma}(k)}{\widetilde{\mathrm{\Delta}}(k)}\Big|_{k=\rmi\kappa_n}=0,
\end{equation}
$\kappa_n=2m|E_n|/\hbar^2$, $E_n$ is the energy of the $n$-th bound state of the electron in potential $V(x)$. To explain it, let us note that the denominator of expression~(\ref{eq:transmission amplitude for arbitrary potential}) is equal to $\det \mathbf{\Gamma}/\widetilde{\mathrm{\Delta}}(k)$. Therefore, the solutions of~(\ref{eq:discrete spectra and gamma matrix}) correspond to the poles of the transmission amplitude~(\ref{eq:transmission amplitude for arbitrary potential}).

Thus, the proposed technique allows to resolve the scattering problem in terms of two linearly independent solutions of the Schr\"{o}dinger equation for potential $V(x)$ and to calculate the binding energies of this potential.

\section{\label{app:parallel}Scattering by the system of parallel quantum wells}
\begin{figure}[h]
\begin{center}
\resizebox{0.3\textwidth}{!}{%
  \includegraphics{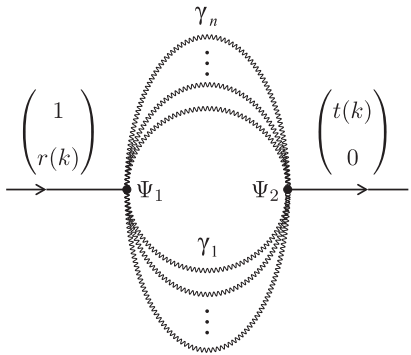}
}
\caption{The graph which corresponds to the scattering problem on the system of $n$ connected in parallel quantum wells. Wavy line represents the potential (in the considered case the potential is the rectangular quantum well).}
\label{fig:parallel quantum wells}
\end{center}
\end{figure}

The problem of electron scattering on the system of $n$ connected in parallel quantum wells (see fig.~\ref{fig:parallel quantum wells}) can be easily solved using the \emph{vertex amplitudes} method.
To do this one need to construct the $\mathbf{\Gamma}$-matrix of the system. In the considered case the elements of $\mathbf{\Gamma}$-matrix have the following form:
$$
\begin{array}{l}
\mathbf{\Gamma}_{11}=1+\displaystyle{\sum\limits_{s=1}^n\GMoel{s}{11}-\GMoel{s}{21}},\\\\
\mathbf{\Gamma}_{12}=\displaystyle{\sum\limits_{s=1}^n\GMoel{s}{12}-\GMoel{s}{22}},\\\\
\mathbf{\Gamma}_{21}=\displaystyle{\sum\limits_{s=1}^n\GMtel{s}{11}\exp(\rmi q_sl_{\gamma_s})+\GMtel{s}{21}\exp(-\rmi q_sl_{\gamma_s})},\\\\
\mathbf{\Gamma}_{22}=\displaystyle{\sum\limits_{s=1}^n\GMtel{s}{12}\exp(\rmi q_sl_{\gamma_s})+\GMtel{s}{22}\exp(-\rmi q_sl_{\gamma_s})}.
\end{array}
$$
Here $q_s=\sqrt{k^2-V_s}$, $V_s=2m_e U_s/\hbar^2$ and $U_s$ is the well's depth on the edge $s$. Let us transform to dimensionless variables by taking $1\,\mbox{nm}$ as length's unit. As one can see, the above $\mathbf{\Gamma}$-matrix coincides with the $\mathbf{\Gamma}$-matrix of the quantum ring up to the replacement of $q$ by $k$. If all wells have a similar depth $U_0$ and width $l$, then the expression for $\mathbf{\Gamma}$-matrix can be significantly simplified. Using~(\ref{eq:gamma matrices differences}), one obtain:
\begin{equation}\label{eq:gamma matrix of parallel quantum wells}
\mathbf{\Gamma}=\left(\begin{array}{cc}
1+\rmi\displaystyle{\frac{n q}{k}}\,\cot ql & -\rmi \displaystyle{\frac{n q}{k}}\, \csc ql \\\\
\rmi \displaystyle{\frac{nq}{k}}\, \csc ql  & -1 - \rmi \displaystyle{\frac{nq}{k}}\, \cot ql
\end{array}\right),
\end{equation}
where $q=\sqrt{k^2-V_0}$.
The transmission and reflection amplitudes can be found using equations~(\ref{eq:amplitudes for two-terminal system}). Namely, for the transmission amplitude we get:
\begin{equation}\label{eq:transmission amplitude parallel}
t_n(k)=\frac{2\rmi n k q}{\left(k^2+n^2q^2\right)\sin ql + 2\rmi nkq\cos ql}.
\end{equation}
As it should be, at $n=1$ this expression turns into the standard expression for the scattering amplitude for the one-dimensional single quantum well. Using~(\ref{eq:transmission amplitude parallel}) we can analyze the dependence of the energy spectrum and transmission coefficient of such a system on the number of connected quantum wells.

At first, let us consider the transmission probability.
\begin{figure}[h]
\begin{center}
\resizebox{0.5\textwidth}{!}{%
  \includegraphics{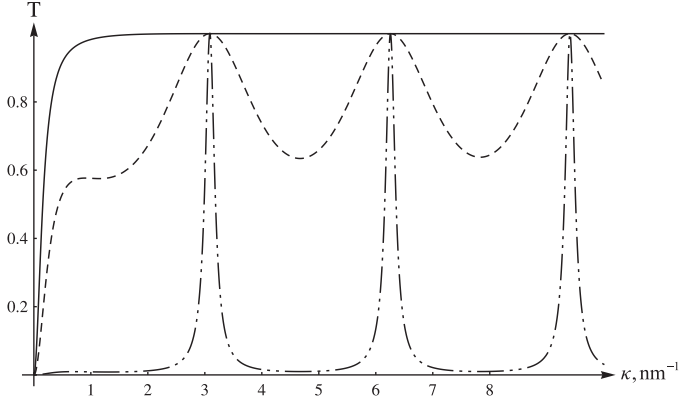}
}
\caption{Transmission coefficient vs. wave number of electron $k$. Solid line corresponds to the single well, dashed line corresponds to the two wells, dashed-dotted line corresponds to the twenty wells. All quantum wells have the same depth and width, namely $U_0=-0.5\, \mbox{eV}$ and $l=1\,\mbox{nm}$. The single quantum well with of such depth and width possesses one bound state with energy equal to $-0.04\, \mbox{eV}$.}
\label{fig:parallel transmission}
\end{center}
\end{figure}
From fig.~\ref{fig:parallel transmission} it follows, for the connected in parallel quantum wells the characteristic oscillations of the transmission coefficient take place. These oscillations are resulted from quantum interference effects in  analogous to the above considered quantum rings. It is also seen that in the region of energy far enough from the well's bottom, $\kappa\gg\sqrt{-V_0}$, the period of oscillations does not depend on the number of wells. Indeed, the elementary calculations show that in this case the transmission coefficient becomes unity at $k=\pi s/l$, $s=1,2,\ldots$ and minimal at $k=(2s+1)\pi/l$, $s=0,1,2,\ldots$. The minimum value $\mathrm{T}_{min}$ is approximately determined by the following expression:
$$
\mathrm{T}_{min}\approx\displaystyle{\frac{4n^2}{1+4n^2+n^4}}.
$$
This expression shows that with increase of the the number of parallel quantum wells the resonances of the transmission coefficient become ``narrower'' and values of transmission coefficient between resonances become closer to zero.

\begin{figure}[h]
\begin{center}
\resizebox{0.49\textwidth}{!}{%
  \includegraphics{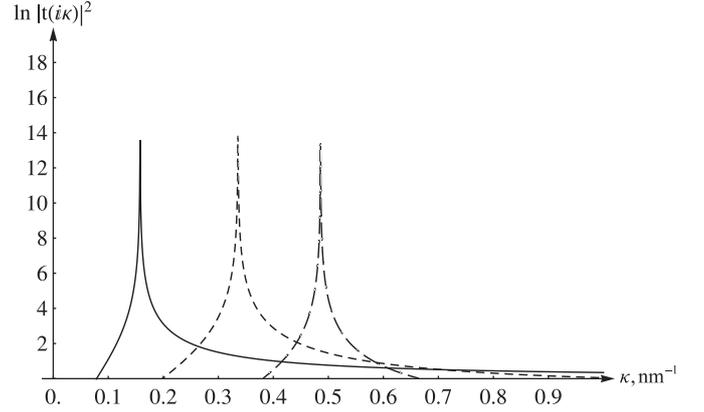}
}
\caption{The logarithm of the transmission coefficient as the function of the number of identical parallel quantum wells. Addition of the new quantum well leads to the shift of the bound state energy position closer to a limiting value $\kappa=\sqrt{-V_0}$. Solid line corresponds to single well, dashed line corresponds to three wells, dashed-dotted line corresponds to ten wells. The single well parameters are $U_0=-0.5\, \mbox{eV}$ and $l=1\,\mbox{nm}$.}
\label{fig:logT}
\end{center}
\end{figure}

\begin{figure}[h]
\begin{center}
\resizebox{0.48\textwidth}{!}{%
  \includegraphics{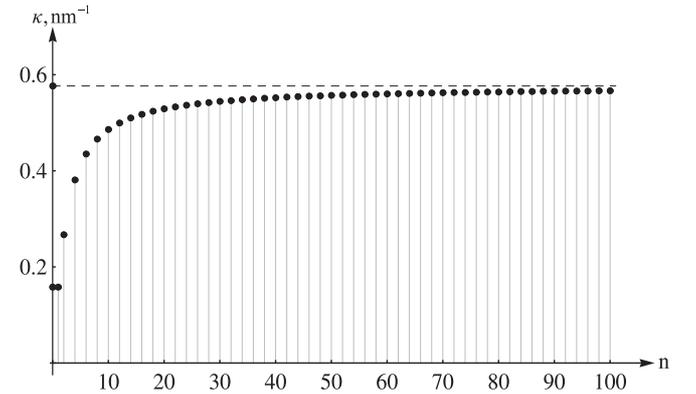}
}
\caption{The dependence of the bound state energy in a system of parallel coupled quantum wells on the number of wells $n$. The dashed horizontal line shows the limiting value $\kappa=\sqrt{-V_0}$. The parameters of wells are $U_0=-0.5\, \mbox{eV}$ and $l=1\,\mbox{nm}$. The parameters of the wells are chosen so that each of them contains the one bound state only with energy equal to $-0.04\, \mbox{eV}$. Points in the vertical axis indicate the value of $\kappa$ for the bound state in single isolated quantum well (lower point) and the value of $\kappa=\sqrt{-V_0}$ (upper point).}
\label{fig:parallel energy shift}
\end{center}
\end{figure}

Using~(\ref{eq:transmission amplitude parallel}) we can analyze the energy spectrum of the system under consideration. To do this, recall that energies of bound states correspond to the simple poles of the transmission amplitude~\cite{LandauIII,Novikov84}. In order to determine the bound states energies construct the analytic continuation of the transmission amplitude $t(\rmi\kappa)$. The bound states are placed in the half-line $\mathrm{Im}\, k>0$. In fig.~\ref{fig:logT} the maxima of the logarithm of the transmission coefficient square correspond to the bound states of the system of $n$ connected in parallel quantum wells. With increase of the number of the quantum number the bound state energy increases and tends to a limiting value $\kappa=\sqrt{-V_0}$, fig.~\ref{fig:parallel energy shift}.

It is interesting to note that if one change the number of connected in parallel identical quantum wells the new bound states in the system will not appear.

%\bibliography{Graph}

\end{document}